\documentstyle[12pt,epsfig,a4]{article} 
\def\bild#1#2{    
        \vspace*{-5mm}
        \begin{center}
        \begin{math}
        \epsfxsize#2cm
        \epsffile{#1}
        \end{math}
        \end{center}  }
\newcommand{\vs}{\vspace{-0.25cm}}
\begin{document} 

\begin{center}
\large{\bf Electromagnetic corrections to the dominant two-pion exchange 
nucleon-nucleon potential}

\medskip

N. Kaiser\\

\smallskip

{\small Physik Department T39, Technische Universit\"{a}t M\"{u}nchen,
    D-85747 Garching, Germany}

\end{center}

\medskip

\begin{abstract}
We calculate at two-loop order in chiral perturbation theory the 
electromagnetic corrections to the dominant two-pion exchange nucleon-nucleon
interaction that is generated by the isoscalar $\pi N$ contact-vertex 
proportional to the large low-energy constant $c_3$. We find that the 
respective $2\pi\gamma$-exchange potential contains sizeable isospin-breaking 
components which amount to about $-1\%$ of the strongly attractive isoscalar 
central $2\pi$-exchange potential. The typical value of these novel  
charge-independence and charge-symmetry breaking central potentials is 
$0.3\,$MeV at a nucleon distance of $r= m_\pi^{-1} = 1.4\,$fm. Our analytical 
result for this presumably dominant $2\pi\gamma$-exchange interaction is in 
a form such that it can be easily implemented into phase-shift analyses and
few-body calculations.            
\end{abstract}



\bigskip
Isospin-violation in the nuclear force is a subject of current interest. 
Significant advances in the understanding of nuclear isospin-violation have 
been made in the past years by employing methods of effective field theory 
(in particular chiral perturbation theory). Van Kolck et al.\,\cite{kolck}
were the first to calculate (in a manifestly gauge-invariant way) the 
complete leading-order pion-photon exchange nucleon-nucleon interaction. In 
addition, the charge-independence and charge-symmetry breaking effects arising
from the pion mass difference $m_{\pi^+}- m_{\pi^0} = 4.59\,$MeV and the
nucleon mass difference $M_n- M_p = 1.29\,$MeV on the (leading order) two-pion
exchange NN-potential have been worked out in Refs.\,\cite{cib,csb}. Recently,
Epelbaum et al.\,\cite{evgenisobr} have continued this line of approach by
deriving the subleading isospin-breaking $2\pi$-exchange NN-potentials and
classifying the relevant isospin-breaking four-nucleon contact terms. 
Moreover, some next-to-leading order corrections to the $\pi\gamma$-exchange
nucleon-nucleon potential (those proportional to the large isovector magnetic 
moment $\kappa_v = 4.7$ of the nucleon) as well as effects from virtual 
$\Delta$-isobar excitation on the $\pi\gamma$-exchange interaction have been 
calculated recently in Ref.\,\cite{pigapot}. 

The long-range (pion-induced) isospin-breaking potentials found so far turned 
out to be rather weak. Typically, their values at a distance of $r=m_\pi^{-1}
=1.4\,$fm lie below $50\,$keV in magnitude (see e.g. Tables I and II in 
Ref.\,\cite{pigapot}). More pronounced and therefore numerically significant 
are the isospin-breaking effects in the subleading $2\pi$-exchange  
interaction (see herefore Figs.\,8 and 10 in Ref.\,\cite{evgenisobr}). As in
the isospin-conserving case this feature can be traced back to the large
magnitude of the low-energy constants $c_2, c_3$ and $c_4$ which enter as
parameters in the second-order chiral pion-nucleon Lagrangian ${\cal L}_{\pi
N}^{(2)}$. According to the numerical investigations made in Ref.\,\cite{
kolck}, the inclusion of the (leading order) $\pi\gamma$-exchange potential 
had negligible effects on the $^1S_0$ low-energy parameters and it led to only 
a tiny  improvement in the fits of the NN-scattering data. With the partial 
cancellations through the next-to-leading order corrections \cite{pigapot} 
taken into account one may then conclude that the long-range 
$\pi\gamma$-exchange interaction is too weak to explain the isospin-breaking
NN-observables. One reason for this feature lies in the relative weakness of
the $1\pi$-exchange interaction itself. For example, its dominant (isovector) 
tensor potential amounts to only $9.2\,$MeV at a distance of $r=m_\pi^{-1} = 
1.4\,$fm. In comparison to this the isoscalar central NN-attraction generated 
by chiral two-pion exchange comes out significantly larger (see e.g. Fig.\,8 in
Ref.\,\cite{nnpot}). Therefore one may expect that electromagnetic corrections
to this strongest long-range NN-interaction, i.e. the chiral $2\pi\gamma
$-exchange interaction, will give rise to more substantial isospin-breaking
effects. It is the purpose of the present short paper to evaluate a selected
class of (presumably dominant) two-loop $2\pi\gamma$-exchange diagrams. Now,
since the isoscalar central NN-attraction comes mainly from the $\pi\pi NN$
contact-vertex proportional to the large low-energy constant $c_3$, one
naturally chooses for a first exploratory study the class of two-loop 
$2\pi\gamma$-exchange diagrams with exactly one such $c_3$-vertex. The
essential long-distance information about the two-loop diagrams is contained 
in their spectral function, which we will evaluate here analytically. 

\bigskip
\medskip

\bild{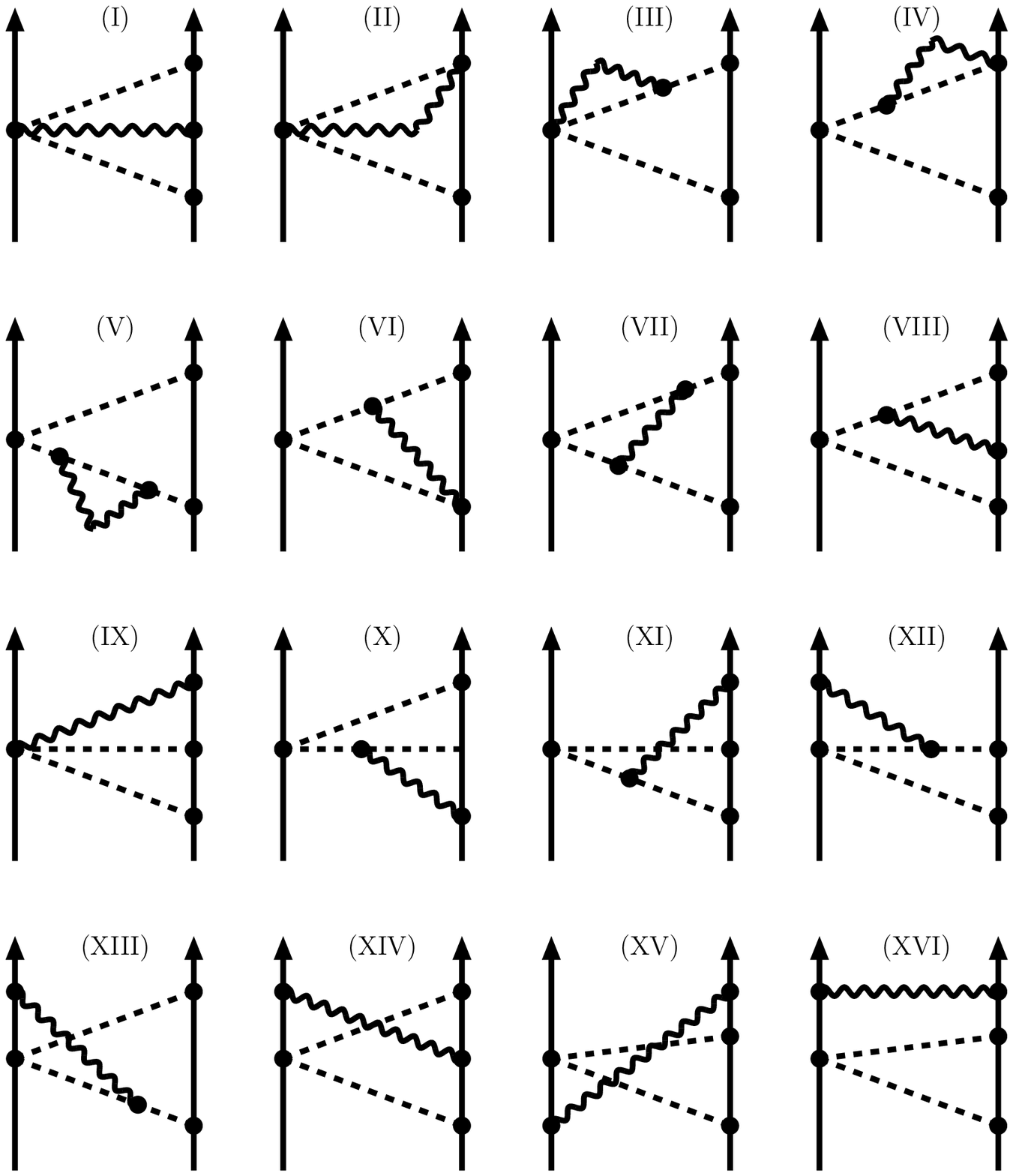}{14}
\vskip -0.2cm
{\it Fig.\,1: Electromagnetic corrections to the dominant isoscalar central 
$2\pi$-exchange NN-interaction generated by the $\pi\pi NN$ contact-vertex
$-2i c_3 f_\pi^{-2}\, \delta^{ab} q_a\cdot q_b$. Diagrams with the 
contact-vertex at the right nucleon line and diagrams turned upside-down are
not shown. The spectral function Im\,$T(i\mu)$ is calculated by cutting the 
intermediate $\pi\pi\gamma$ three-particle state.}

\bigskip

As a result, we do indeed find sizeable charge-independence and
charge-symmetry breaking central potentials of magnitude $0.3\,$MeV at $r=
m_\pi^{-1}=1.4\,$fm  (for $c_3 = -3.3\,$GeV$^{-1}$). These seem to be the 
largest long-range isospin-breaking NN-potentials obtained so far. In order to
test their phenomenological relevance they should be implemented into future
phase-shift  analyses and few-body calculations. As a result, we do indeed
find  sizeable charge-independence and
charge-symmetry breaking central potentials of magnitude $0.3\,$MeV at $r=
m_\pi^{-1}=1.4\,$fm (for $c_3 = -3.3\,$GeV$^{-1}$). These seem to be the
largest long-range isospin-breaking NN-potentials obtained so far. In order to
test their phenomenological relevance they should be implemented into future
phase-shift analyses and few-body calculations.  

Let us start with recalling the chiral $2\pi$-exchange NN-interaction. Its
dominant contribution to the isoscalar central channel comes from a triangle
diagram involving one $\pi\pi NN$ contact-vertex with momentum-dependent 
coupling: $-2i c_3 f_\pi^{-2}\,\delta^{ab} q_a\cdot q_b$. The corresponding 
potential in coordinate space (obtained by a Laplace transformation) reads 
\cite{nnpot}:
\begin{equation} \widetilde V_C^{(2\pi)}(r)={3c_3 g_A^2 \over 32\pi^2 f_\pi^4}
\, {e^{-2z} \over r^6} \,(6+12z+10z^2+4z^3+z^4) \,, \end{equation}
where $z= m_\pi r$. The occurring parameters are: $g_A= 1.3$ (the nucleon 
axial-vector coupling constant), $f_\pi= 92.4\,$MeV (the pion decay constant),
$m_\pi=139.57\,$MeV (the charged pion mass) and the low-energy constant $c_3 = 
-3.3\,$GeV$^{-1}$. The latter value is the average of $c_3 = -3.2\,
$GeV$^{-1}$ and  $c_3=-3.4\,$GeV$^{-1}$ obtained by Entem et al.\,\cite{entem}
and Epelbaum et al.\,\cite{evgeni1,evgeni2}, respectively, in fits to 
NN-scattering phase shifts at next-to-next-to-next-to-leading order in the 
chiral expansion. The numerical values in the first row of Table\,I display the
magnitude and $r$-dependence of this dominant isoscalar central $2\pi$-exchange
potential $\widetilde V_C^{(2\pi)}(r)$.  

Now we will add to the $2\pi$-exchange triangle diagram a photon line which
runs from one side to the other. There are five positions for the photon to
start on the left hand side and seven positions to arrive at the right hand
side. Leaving out those four diagrams which vanish in Feynman gauge (with
photon propagator proportional to $g_{\mu\nu}$) we get the 16 representative
diagrams shown in Fig.\,1. Except for diagram (I) these are to be understood  
as being duplicated by horizontally reflected partners. A further doubling of
the number of diagrams comes from interchanging the role of both nucleons. 

We are interested here in the coordinate-space potentials generated by 
the $2\pi\gamma$-exchange diagrams shown in Fig.\,1. For that purpose it is 
sufficient to know their spectral function or imaginary part Im\,$T(i\mu)$. 
Making use of (perturbative) unitarity in the form of the Cutkosky cutting 
rule we can calculate the two-loop spectral functions as integrals of the
(subthreshold) $\bar NN \to \pi\pi \gamma \to \bar NN $ transition amplitudes 
over the Lorentz-invariant $2\pi\gamma$ three-particle phase space
\cite{2looppot}. In the (conveniently chosen) center-of-mass frame this leads 
to two angular integrations and two integrals over the pion energies. Due to 
the heavy nucleon limit ($M_N \to \infty$) and the masslessness of the
photon ($m_\gamma= 0$) several simplifications occur and therefore most of 
these integrations can actually be performed in closed analytical form. For a
concise presentation of our results it is also advantageous to scale out all 
common (dimensionful) parameters from the spectral function:
\begin{equation} {\rm Im}\, T(i\mu) = {\alpha c_3 g_A^2 m_\pi^3 \over \pi (4
f_\pi)^4} \, S\Big({\mu \over m_\pi} \Big) \,, \end{equation}
and to work with the dimensionless variable $u = \mu/m_\pi$, where $\mu\geq
2m_\pi$ denotes the $\pi\pi\gamma$-invariant mass and $\alpha = 1/137.036$ is
the fine-structure constant. Without going into further technical 
details, we enumerate now the contributions of the 16 representative
$2\pi\gamma$-exchange diagrams shown in Fig.\,1 to the dimensionless spectral
function $S(u)$. We find for $u\geq 2$:    
\begin{equation} S(u)^{(\rm I)}=(2-\tau_1^3 -\tau_2^3) \bigg\{ {u^3 \over 2}-5u
+4 +{2 \over u-1} +{2\over u} \ln(u-1) \bigg\} \,, \end{equation} 
where $\tau_{1,2}^3$ denotes the third components of the usual isospin
operators. 
\begin{equation} S(u)^{(\rm II)} = 36u -2u^3 -48  -{8 \over u-1} -{40\over u} 
\ln(u-1) \,, \end{equation} 
\begin{equation} S(u)^{(\rm III)} = 9u^3 -12u^2 -54u +84  +{4\over
u}(14u^2-11-3u^4)  \ln(u-1) \,, \end{equation} 
\begin{equation} S(u)^{(\rm IV)} = 7u^3 -12u^2 -26u +44  +{4\over
u}(8u^2-5-3u^4)  \ln(u-1) \,, \end{equation} 
\begin{equation} S(u)^{(\rm V)} = 16u^2 -12u^3 +56u -80  -{16\over u}(u^2-1)^2
\ln(u-1) 
\end{equation} 
\begin{equation} S(u)^{(\rm VI)}= 3u^3 -8u^2 -22u +48 +{4 \over u-1} +{44 \over
u} \ln(u-1) +{16 \over u}(u^2-2) \int_1^{u/ 2}\! {dx \over y} \, \ln{u(x+y)-1 
\over u(x-y)-1} \,, \end{equation} 
with the abbreviation $y = \sqrt{x^2-1}$. 
\begin{eqnarray} S(u)^{(\rm VII)}&=& 20u^2-9u^3 +22u -52+{4 \over u}(5u^4-16u^2
+11) \ln(u-1) \nonumber \\ && +{8 \over u^2}(u^2-2) \oint_1^{u/2}\!\! dx \, 
{(ux -1) (4-u^2-2u x)\over (u-2x) y}\, \ln{u(x+y)-1 \over u(x-y)-1} \,,
\end{eqnarray}  
\begin{eqnarray} S(u)^{(\rm VIII)}&=&(2-\tau_1^3-\tau_2^3)\bigg\{ 3u^2 +6u -11 
+\bigg(u^3-2u -{2\over u} \bigg)  \ln(u-1) \nonumber \\ &&  -{3u^3 \over 2} 
-{1 \over u-1}+{2 \over u}(u^2-2) \int_1^{u/2}\!\! dx \, {ux -1 \over y}\, 
\ln{u(x+y)-1 \over u(x-y)-1}\bigg\} \,, \end{eqnarray} 
\begin{eqnarray} S(u)^{(\rm IX)}&=& (2+\tau_1^3+\tau_2^3) \bigg\{ {u^3 \over 2}
-5u + 4+ {2 \over u-1}+{2\over u}  \ln(u-1) \nonumber \\ &&  +{1\over 3}(26
-5u^2) \sqrt{u^2-4} -{8  \over u} \ln{ u +\sqrt{u^2-4} \over 2} \bigg\}  \,, 
\end{eqnarray} 
\begin{eqnarray} S(u)^{(\rm X)}&=&(2+\tau_1^3+\tau_2^3)\bigg\{ u^2-{u^3\over 4}
+{3 u \over 2}-5 +\bigg(u^3-4u + {3\over u}\bigg)\ln(u-1) \nonumber \\ &&  +{8
\over 3}(1-u^2) \sqrt{u^2-4} +\bigg( 4u^3 -10u +{8  \over u}\bigg) \ln{ u +
\sqrt{u^2-4} \over 2} \nonumber \\ && + {4 \over u}(u^2-2) \oint_1^{u/2} \!\!\!
{dx \over u-2x}\bigg[u y+(1-u^2) \ln{u-x+y \over u-x-y } \bigg] \bigg\}  \,,
\end{eqnarray}  
\begin{eqnarray} S(u)^{(\rm XI)}&=&S(u)^{(\rm X)}+(2+\tau_1^3+\tau_2^3)\bigg\{ 
u^2-u^3 +\bigg(6u -u^3- {8\over u}\bigg)\ln(u-1) \nonumber \\ &&  +3u-1-
{1 \over u-1} +{2 \over u}(u^2-2) \int_1^{u/2}\!\!\! dx \, {ux -1 \over y}\, 
\ln{u(x+y)-1 \over u(x-y)-1}\bigg\}  \,, \end{eqnarray} 
\begin{eqnarray} S(u)^{(\rm XII)} &=& (\tau_1^3+\tau_2^3+2\tau_1^3\tau_2^3) 
\bigg\{ {u^3\over 4}-u^2- {3 u \over 2}+5 +\bigg(4u -u^3- {3\over u}\bigg)
\ln(u-1) \nonumber \\ && - 3u^2 \sqrt{u^2-4} +\bigg( 4u^3 -6u +{8  \over u}
\bigg) \ln{ u +\sqrt{u^2-4} \over 2} \nonumber \\ && + {4 \over u}(u^2-2) 
\oint_1^{u/2} \!\!\! {dx \over u-2x}\bigg[u y+(1-u^2) \ln{u-x+y \over u-x-y }
\bigg] \bigg\}  \,, \end{eqnarray} 
\begin{equation} S(u)^{(\rm XIII)} =  S(u)^{(\rm XII)} +(\tau_1^3+\tau_2^3+2
\tau_1^3 \tau_2^3)  \bigg\{ 2u^2-{u^3\over 2}+3 u -10 +\bigg(2u^3-8u+ {6\over
u}\bigg) \ln(u-1) \bigg\} \,, \end{equation} 
\begin{equation} S(u)^{(\rm XIV)} = (3+\tau_1^3+\tau_2^3-\tau_1^3 \tau_2^3)  
\bigg\{ \bigg({u^2 \over 2}-3\bigg) \sqrt{u^2-4} +{4\over u}\ln{ u +
\sqrt{u^2-4} \over 2} \bigg\} \,, \end{equation}  
\begin{equation} S(u)^{(\rm XV)} + S(u)^{(\rm XVI)} = 0  \,. \end{equation} 
The exact cancellation between diagram (XV) and the irreducible part of diagram
(XVI) has the same kinematical reason as the cancellation between the crossed 
box and (the irreducible part of the) planar box diagram, which has been 
discussed in detail in Ref.\,\cite{nnpot}. We have carefully checked 
gauge-invariance. The total (dimensionless) spectral function $S(u)$ stays 
$\xi$-independent when adding a longitudinal part to the photon propagator: 
$g_{\mu\nu} \to g_{\mu\nu}+ \xi\, k_\mu k_\nu$. This property holds already
separately in the subclasses of diagrams: (I)+(VIII), (II)+(IV)+(VI), 
(III)+(V)+(VII), (IX)+(X)+(XI), (XII)+(XIII) and (XIV). The ''encircled'' 
integrals appearing in Eqs.(9,12,14) involve the following regularization
prescription:  
\begin{equation} \oint_1^{u/2}\!\!dx\, {f(x)\over u-2x} =  \int_1^{u/2} \!\! 
dx\, { f(x)-f(u/2) \over u-2x} \,. \end{equation}
This regularization prescription eliminates from some diagrammatic 
contributions ((V), (VII), (X), (XI), (XII), (XIII)) to the spectral function 
an infrared singularity arising from the emission of soft photons ($\bar N
N \to \pi\pi\gamma_{\rm soft}$). In the $2\pi \gamma$ phase space (an
oval-shaped region in the $\omega_1\omega_2$-plane spanned by the pion 
energies) the singularity is located at that extremal boundary point
($\omega_{1,2}= \mu/2$) where the photon energy $\mu -\omega_1-\omega_2$
becomes identical to zero. The singular factor $(u-2x)^{-1}$ in Eq.(18) stems
from a pion propagator. The regularization prescription defined in Eq.(18) is
equivalent to the familiar ''plus''-prescription \cite{peskin} employed
commonly for parton splitting functions in order to eliminate there an
analogous infrared singularity due to soft gluon radiation.\footnote{It would
of course be desirable to extend the present  calculational framework (by
including various radiation diagrams etc.) such that the overall infrared
finiteness could be demonstrated in detail. This goes beyond the scope of the
present work and therefore we stay with the physically well-founded
regularization prescription Eq.(18).} We note as an aside that  the
non-elementary integrals ($\int_1^{u/2} dx \dots$) in Eqs.(8--14) can be solved
in terms of dilogarithms and squared logarithms of the argument
$(u+\sqrt{u^2-4})/2$ \cite{formulas}.

Let us  first discuss  some generic properties. All contributions to the
dimensionless spectral function $S(u)$ vanish at the $\pi\pi\gamma$-threshold 
$u= 2$. This is markedly different from the original one-loop two-pion 
exchange triangle diagrams which possess the discontinuous spectral function 
$S(u)^{(2\pi)}=- 12\pi (\alpha u)^{-1}(u^2-2)^2 \, \theta(u-2)$ with a
non-vanishing threshold value. The leading threshold behavior of the total 
dimensionless spectral function $S(u)$ is: $S(u)=64(1+\tau_1^3)(1+\tau_2^3)(1-
\ln2) \sqrt{u-2}+{\cal O}(u-2)$ and it comes exclusively from the four 
diagrams  (X)--(XIII). Such a square-root like growth implies an asymptotic 
tail of the coordinate-space potential of the form $e^{-2m_\pi r} r^{-5/2}$ 
for $r\to \infty$, whereas at short distances $r\to 0$ the potential develops 
a $r^{-6}\ln^2(m_\pi r)$ singularity, somewhat stronger than that of 
$\widetilde V_C^{(2\pi)}(r)$ written in Eq.(1). At any intermediate nucleon 
distance $r$ the $2\pi\gamma$-exchange central potential in coordinate-space 
$\widetilde V_C^{(2\pi\gamma)}(r)$ can be calculated from the spectral
function Im\,$T(i\mu)$ via a modified Laplace transformation:
\begin{eqnarray} \widetilde V_C^{(2\pi\gamma)}(r) &=& - {1\over 2\pi^2 r}
\int_{2m_\pi}^\infty \!d\mu\, \mu e^{-\mu r} \, {\rm Im} \,T(i\mu) \nonumber
\\ &=& \widetilde V_C^{(0)}(r)  + \tau_1^3 \tau_2^3\, \widetilde
V_C^{(\rm cib)}(r) +(\tau_1^3+ \tau_2^3)\, \widetilde V_C^{(\rm csb)}(r)\,, 
\end{eqnarray}
where we have given in the second line of Eq.(19) its decomposition into
isospin-conserving (0), charge-independence breaking (cib), and charge-symmetry
breaking (csb) parts. We note again as an aside that the (partial) 
contributions to the potential $\widetilde V_C^{(2\pi\gamma)}(r)$ coming from 
the terms in Eqs.(3--16) without integrals can be represented by the 
exponential function $e^{-2z}$, the exponential-integral function
$e^{-z}E_1(z)$, and two modified Bessel functions $K_{0,1}(2z)$, each
multiplied by a pure polynomial in $1/z$ of degree 6 or lower, where $z= m_\pi 
r$ \cite{formulas}.

We are now in the position to present numerical results for the dominant
$2\pi\gamma$-exchange nucleon-nucleon potentials. The numbers in the second, 
third, and fourth row of Table\,I show the dropping of the central potentials
$\widetilde V_C^{(0)}(r)$, $\widetilde V_C^{(\rm cib)}(r)$ and $\widetilde 
V_C^{(\rm csb)}(r)$ with the nucleon distance $r$ in the range $1.0\,{\rm fm}
\leq r \leq 2.1\,{\rm fm}$. Note that all potentials are given in units of
MeV. One observes that the repulsive charge-independence breaking (cib)
potential and the repulsive charge-symmetry breaking (csb) potential are 
approximately equal and about half as strong as the isospin-conserving one. 
Their magnitudes (for example, $0.3\,$MeV at a distance of $r=1.4\,$fm) are 
exceptionally large in comparison to all so far known isospin-breaking 
NN-potentials generated by long-range pion-exchange. The dominant 
contributions to $\widetilde V_C^{(\rm cib)}(r)$ and $\widetilde V_C^{(\rm 
csb)}(r)$ come from the four diagrams (X)--(XIII). This selective feature is 
consistent with the above-mentioned leading threshold behavior of the 
spectral function $S(u)$ and the associated large-$r$ asymptotics.

\vskip -0.3cm
\begin{table}[hbt]
\begin{center}
\begin{tabular}{|c|cccccccccccc |}
\hline 
$r$~[fm]& 1.0 & 1.1 & 1.2 & 1.3 & 1.4 & 1.5 & 1.6 & 1.7 & 1.8 & 1.9 & 2.0&2.1 
\\ \hline
$\widetilde V_C^{(2\pi)}$ &--221 &--121 & --69.3 &--41.4 & --25.6 &--16.3
   & --10.6 &--7.05 &--4.78 & --3.30 & --2.31 & --1.64\\
$\widetilde V_C^{(0)}$ & 6.47 & 3.25 & 1.72 & 0.953 & 0.547 & 0.325 & 0.198 & 
0.124 & 0.079 & 0.051 & 0.034 & 0.023 \\
$\widetilde V_C^{(\rm cib)}$ & 3.54 & 1.78 & 0.948 & 0.526 & 0.303 & 0.180 &
0.110 & 0.069 & 0.044 & 0.029 & 0.019 & 0.013 \\ 
    $\widetilde V_C^{(\rm csb )}$& 3.57 & 1.80 & 0.956 & 0.530 & 0.305 & 
0.182 & 0.111 & 0.070 & 0.045 & 0.029  & 0.019 & 0.013\\ \hline
\end{tabular}
\end{center}
\vskip -0.2cm
{\it Table I: The dominant isoscalar central $2\pi$-exchange NN-potential 
$\widetilde V_C^{(2\pi)}(r)$, and electromagnetic corrections to it, as a 
function of the nucleon distance $r$. The values in the third and fourth row
correspond to the isospin-violating potentials  $\widetilde V_C^{(\rm
cib)}(r)$ and $\widetilde V_C^{(\rm csb)}(r)$. All potentials are given in 
units of MeV.}      
\end{table}
\vskip -0.2cm

It is also instructive to compare the isospin-violating
$2\pi\gamma$-exchange potentials with the isospin-conserving $2\pi$-exchange
potential $\widetilde V_C^{(2\pi)}(r)$. Their relative ratio of about 
$-1$ percent is consistent with the usual rule of thumb estimate, namely 
$\alpha/\pi \simeq 1/430$ times a numerical factor. In the present case this
numerical factor is not just 1 but of the order 5, due to the large number of 
contributing diagrams (in total there were 54 non-vanishing diagrams).       

We have also computed the $2\pi\gamma$-exchange central potentials 
proportional to the other large low-energy constant $c_2 = 
3.3\,$GeV$^{-1}$. At a reference distance of $r_0=1.4\,$fm one finds the values
 $\widetilde V_C^{(0)}(r_0)= -8.54 \,$keV, $\widetilde V_C^{(\rm cib)}(r_0)= 
-23.4\,$keV and $\widetilde V_C^{(\rm csb)}(r_0)=-14.8\,$keV. These potentials 
are in fact more  than an order of magnitude smaller than their counterparts 
generated by the $c_3$-vertex. Furthermore, as expected the $2\pi\gamma
$-exchange central potentials arising from the $c_1$-vertex 
are small. With $c_1 = -0.8\,$GeV$^{-1}$ we get $\widetilde V_C^{(0)}(r_0)= 
23.4 \,$keV, $\widetilde V_C^{(\rm cib)}(r_0)= 8.40\,$keV and $\widetilde
V_C^{(\rm csb)}(r_0)=9.78\,$keV. This confirms our assumption made about the
dominant $2\pi\gamma$-exchange NN-potential.   
 
In summary we have calculated in this work the electromagnetic corrections to 
the dominant two-pion exchange nucleon-nucleon interaction that is generated 
by the isoscalar $\pi N$ contact-vertex proportional to the large low-energy 
constant $c_3=-3.3\,$GeV$^{-1}$. We have evaluated analytically the spectral 
function of a large class of 70 two-loop $2\pi\gamma$-exchange diagrams. The
result for the total spectral function is manifestly gauge-invariant. An 
infrared singularity related to soft photon emission which occurs in some
contributions has been regularized by a prescription analogous to the one
employed commonly for parton splitting functions. As a major result we have
found that the $2\pi\gamma$-exchange potential contains sizeable
isospin-breaking components which amount to about $-1\%$ of the strongly
attractive isoscalar central $2\pi$-exchange potential. A typical value of the
charge-independence and charge-symmetry breaking central potential is
$0.3\,$MeV at a nucleon distance of $r= m_\pi^{-1} =1.4\,$fm. Our analytical
result for this novel and exceptionally large isospin-violating long-range
nuclear force has been presented in a form such that it can be easily
implemented into phase-shift analyses and few-body calculations. For
completeness one should also calculate the $2\pi\gamma$-exchange diagrams 
proportional to the remaining large low-energy constant $c_4 \simeq 3.4
\,$GeV$^{-1}$ which generate (isospin-violating) spin-spin and tensor 
potentials. Some of the technical advantages arising from the Lorentz-scalar 
nature of the $c_3$-vertex are then, however, no more available. The issue of 
(overall) infrared finiteness should also be further explored. Work along this
line is in progress.


\begin{thebibliography}{99}
\bibitem{formulas} These somewhat lengthy formulas can be obtained from the 
author upon request. The information given here in Eqs.(2-19) is already 
sufficient for an easy and accurate numerical evaluation of the potential
$\widetilde V_C^{(2\pi\gamma)}(r)$.\vs
\end{thebibliography}
\end{document}